\documentclass[12pt,preprint]{aastex}
\begin{document}

\title{Identification of the Mass Donor Star's Spectrum in SS~433}
\author{T. C. Hillwig\altaffilmark{1}, D. R. Gies\altaffilmark{1,2}, W. Huang\altaffilmark{2}, M. V. McSwain}
\affil{Center for High Angular Resolution Astronomy, Department of Physics and Astronomy, Georgia State University, Atlanta, GA 30303}
\affil{thillwig@chara.gsu.edu, gies@chara.gsu.edu, huang@chara.gsu.edu, mcswain@chara.gsu.edu}
\author{M. A. Stark\altaffilmark{2}}
\affil{Department of Astronomy and Astrophysics, The Pennsylvania State University, 525 Davey Laboratory, University Park, PA 16802}
\affil{stark@astro.psu.edu}
\author{A. van der Meer, and L. Kaper}
\affil{Astronomical Institute ``Anton Pannekoek'', University of Amsterdam, Kruislaan 403, NL-1098 SJ Amsterdam, Netherlands}
\altaffiltext{1}{Visiting Astronomer, Kitt Peak National Observatory, 
National Optical Astronomical Observatories, which is operated by the 
Association of Universities for Research in Astronomy, Inc. (AURA) 
under cooperative agreement with the National Science Foundation.}
\altaffiltext{2}{Guest Observer, McDonald Observatory of the University
of Texas at Austin.}
\affil{ameer@science.uva.nl, lexk@science.uva.nl}

\begin{abstract}

We present spectroscopy of the microquasar SS~433 obtained near primary
eclipse and disk precessional phase $\Psi = 0.0$, when the accretion
disk is expected to be most ``face-on''.  The likelihood of
observing the spectrum of the mass donor is maximized at this combination of
orbital and precessional phases since the donor is in the foreground and
above the extended disk believed to be present in the system.  The spectra
were obtained over four different runs centered on these special phases.
The blue spectra show clear evidence of absorption features
consistent with a classification of A3-7~I.
The behavior of the observed lines indicates an
origin in the mass donor.  The observed radial velocity variations
are in anti-phase to the disk, the absorption lines strengthen
at mid-eclipse when the donor star is expected
to contribute its maximum percentage of the total flux, and the line
widths are consistent with lines created in an A supergiant photosphere.
We discuss and cast doubt on the possibility that these lines
represent a shell spectrum
rather than the mass donor itself.
We re-evaluate the mass ratio of the system
and derive masses of $10.9\pm 3.1~M_\odot$ and $2.9\pm0.7~M_\odot$
for the mass donor and compact object plus disk, respectively.  We
suggest that the compact object is a low mass black hole.

In addition, we review the behavior of the observed emission lines from
both the disk/wind and high velocity jets.

\end{abstract}

\section{INTRODUCTION}

The well-studied binary SS~433, despite the wealth of observational
and theoretical studies, is still one of the most enigmatic systems.
SS~433 is an X-ray binary consisting of a compact star and a mass
donor star, called such
because it donates mass to a precessing extended disk surrounding the
compact companion, typically considered to be a black hole
\citep[e.g.][]{lei84} or neutron star \citep[e.g.][]{dod91}.  SS~433
is also well known as a source of collimated relativistic jets with
speeds $v_{jet}\approx 0.26c$ \citep[e.g.][]{mar82}.  The number
of components in this system combined with their complexity greatly
obscures interpretation of observational data and theoretical
modeling, and thus many physical parameters of the system are
as yet undetermined.

The 13 day orbital period of the binary as well as the 162 day period
of the disk precession are well established \citep{gor98,eik01,gie02}.
The radial velocity behavior of the compact companion and inner
disk, while complicated by emission from other components,
has also been convincingly established
with a semiamplitude of $\approx 170$ km s$^{-1}$\citep{fab90,gie02}.
\citet*{ant87} suggested that observation of the mass
donor spectrum could be possible and would lead to direct kinematical
masses for the two components.  An accurate
detection of the mass donor spectrum is crucial in determining if
the compact companion is a black hole or neutron star.  A first
attempt at this was made by \citet[][= GHM03]{gie03}
who found similarities
between absorption features in the mid-eclipse spectrum of SS~433 and
an A-type evolved star, specifically the A7 Ib star HD~148743.  Their
detection provided a tentative determination of the radial velocity
curve of the mass donor resulting in a mass determination for both
components.  We set out to confirm this tentative detection of the
mass donor with additional observations.

Absorption features from the mass donor star may best
be observed during primary eclipse (donor star inferior
conjunction) when much of the continuum light from the disk is blocked
and the donor star is in the foreground.  The precession of the
extended disk complicates detection of the mass donor since we expect
the disk to have both a high opacity and large vertical extent.
This means that
light from the mass donor will be obscured by the disk except
near precessional phase zero, the phase of maximum disk opening.
The combination of precessional phase $\Psi = 0.0$ and
orbital phase $\phi = 0.0$ limits the possible observing
times to roughly two five-night windows per year.

Here we present new spectroscopy of SS~433 at three epochs corresponding
to $\Psi = 0.0$ and $\phi = 0.0$ that we combine with the original data
from GHM03.  We show that the absorption spectrum
is well matched by the spectrum of HD~9233, an A4~Iab star.  We also
utilize the emission features in the spectrum to confirm both the
orbital and precessional phases of observation.  The observed jet
features and the ``stationary'' emission lines are also discussed.

\section{Observations and Reductions}

The spectra of SS~433 were obtained during four epochs of
observation.  We obtained spectra during 2002 June, 2002 November,
and 2003 April with the Large Cassegrain Spectrograph
on the 2.7m Harlan J. Smith Telescope at the University of
Texas McDonald Observatory.  The
fourth epoch spectra were obtained with the RC Spectrograph
at the Mayall 4m telescope at Kitt
Peak National Observatory in 2003 October.
Two different CCDs, the TI1 and CC1 arrays, were used during
the McDonald Observatory runs.  During the 2002 November run 
we used the CC1 array, which unfortunately produced a low S/N ratio.
The first night of the 2003 April run also used this CCD, but
the remaining nights utilized the TI1 array in a successful
attempt to improve the results.
Table \ref{data} provides a list of the mid-exposure heliocentric
Julian date, observatory of origin, wavelength coverage,
reciprocal dispersion, resolving power, and S/N ratio in the continuum (near
4600 \AA ) for each night of the four observing runs.  The KPNO
observations include wavelengths out to 4940\AA , though
internal vignetting limits quantitative analysis to wavelengths
blueward of about 4800\AA .  The final
two columns provide the calculated precessional ($\Psi$) and orbital
($\phi$) phases
of each observation.  For orbital phase we use the light curve
ephemeris of \citet{gor98},
$$\mathrm{HJD}\ 2,450,023.62+13.08211E$$
and for disk precession the model ephemeris of \citet{gie02},
$$\mathrm{HJD}~2,451,458.12+162.15E.$$
\begin{center}
\begin{deluxetable}{lccccccc}
\tabletypesize{\footnotesize}
\tablewidth{0pc}
\tablecolumns{8}
\tablecaption{Information for the four epochs of observation\label{data}}
\tablehead{
\colhead {Date}        & \colhead {} & \colhead {$\lambda$ range} &
\colhead {Dispersion}  & \colhead {$R$} & \colhead {S/N ratio} &
\colhead {}      & \colhead {}  \\
\colhead {(HJD - 2,450,000)}  & \colhead {Observatory} & \colhead {(\AA )} &
\colhead {(\AA ~pixel$^{-1}$)}  & \colhead {($\lambda/\Delta\lambda)$} &
\colhead {(res. element$^{-1}$)} & \colhead {$\Psi$} & \colhead {$\phi$} }
\startdata \hline
2430.762\dotfill & McDonald & 4060--4750 & 0.889 &2600&\phn31& 0.998 & 0.003 \\
2431.823\dotfill & McDonald & 4060--4750 & 0.889 &2600&\phn16& 0.004 & 0.084 \\
2432.938\dotfill & McDonald & 4060--4750 & 0.889 &2600&\phn40& 0.012 & 0.169 \\\hline
2590.037\dotfill & McDonald & 4035--4745 & 0.694 &3300&\phn\phn9& 0.965 & 0.988 \\\hline
2755.873\dotfill & McDonald & 4070--4790 & 0.703 &3300&\phn42& 0.003 & 0.854 \\
2756.873\dotfill & McDonald & 4060--4760 & 0.887 &2600&\phn66& 0.010 & 0.931 \\
2757.870\dotfill & McDonald & 4060--4760 & 0.887 &2600&\phn24& 0.016 & 0.007 \\
2759.878\dotfill & McDonald & 4060--4760 & 0.887 &2600&\phn60& 0.028 & 0.160 \\\hline
2912.695\dotfill & KPNO    & 4180--4940 & 0.372 &6200& 183 & 0.971 & 0.842 \\
2913.692\dotfill & KPNO    & 4180--4940 & 0.372 &6200& 102 & 0.977 & 0.918 \\
2914.697\dotfill & KPNO    & 4180--4940 & 0.372 &6200&\phn81& 0.983 & 0.995 \\
2915.689\dotfill & KPNO    & 4180--4940 & 0.372 &6200&\phn66& 0.989 & 0.071 \\
\enddata
\end{deluxetable}
\end{center}

The spectra were reduced using standard routines in IRAF\footnotemark.
\footnotetext{IRAF is distributed by the National Optical Astronomical
Observatories, which is operated by the Association of Universities for
Research in Astronomy, Inc., under cooperative agreement with the
National Science Foundation.} 
All the spectra from an individual night were coadded to
improve the S/N ratio.  No phase shifts were introduced during
coaddition since the total observing time each
night ($< 4$ hrs) is negligible relative to the orbital period.
The coadded spectra were then shifted to a heliocentric frame and
rectified to a unit continuum by fitting regions free from
emission lines.  The continuum rectification arbitrarily
removes the continuum variations caused by the eclipse occuring during
the observation intervals.  All of the spectral intensities in
this paper are set relative to the continuum, which changes on a
night-to-night basis.

\section{The Emission Spectrum of SS~433}

\begin{figure}[p]
\begin{center}
\plotone{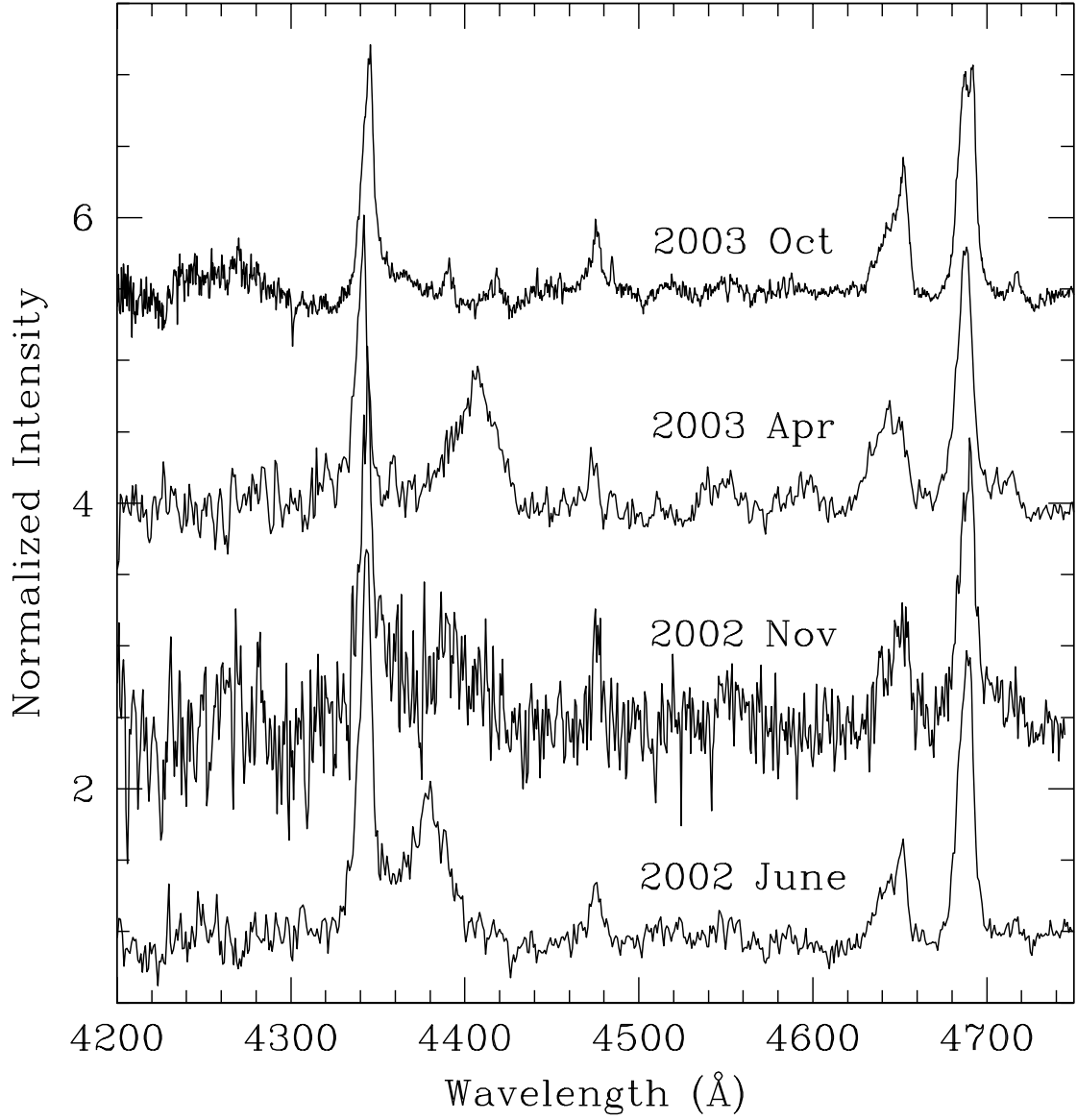}
\end{center}
\caption[SS~433 mid-eclipse spectra] {SS~433 spectra obtained
closest to mid-eclipse for each of the four observing runs.  The
continua are set at 1.0, 2.5, 4.0, and 5.5 in normalized intensity
units for clarity of separation.
\label{mespec}}
\end{figure}

We begin by considering the emission lines that dominate the
appearance of the blue spectrum of SS~433 and that can be used
to verify the orbital and precessional phases of our observations.
Figure \ref{mespec} shows the spectra obtained nearest to
mid-eclipse from
each of the four runs.  The emission features can generally be
grouped into two categories, the ``stationary'' lines,
which presumably originate in or near the disk
\citep{cra80,cra81}, and the relativistic jet lines.  Both sets
of lines typically show variability from night-to-night and
run-to-run in strength,
structure, and velocity.  Here we discuss the variations in strength
and velocity of the emission features in our spectra
for both sets of lines.

\subsection{Collimated Jet Lines}

\begin{figure}[p]
\begin{center}
\plotone{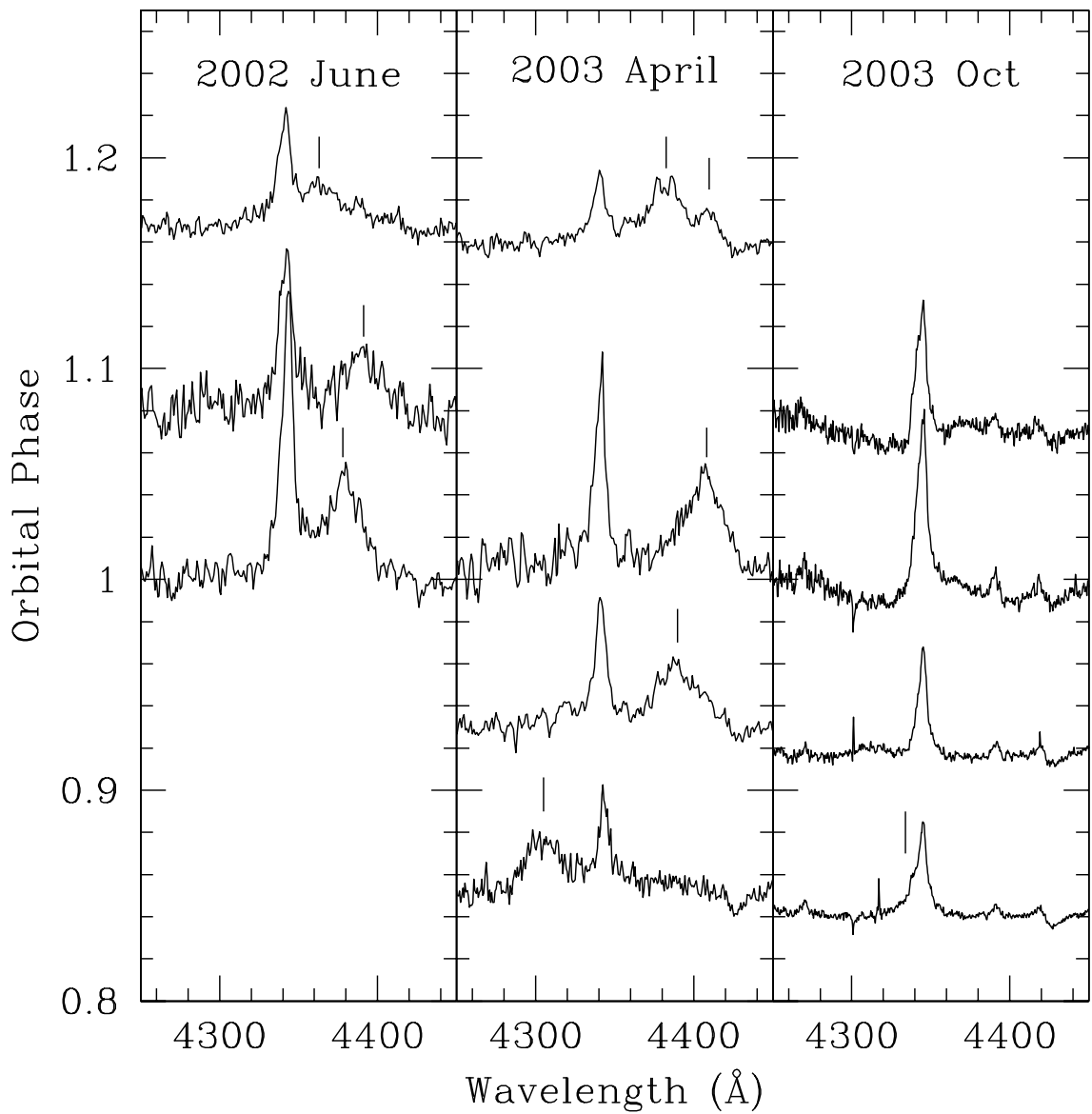}
\end{center}
\caption[Jet line variations] {The variability in velocity and
strength of H$\beta -$ jet lines for the three epochs with more than one
observation.  The spectra are scaled so that their continuum
corresponds to the orbital phase of observation.  Jet line
positions are indicated by vertical tick marks.  H$\gamma$ emission
is detected at 4340 \AA .
\label{jets}}
\end{figure}

Figure \ref{jets} shows the progression of the blue-shifted H$\beta$
jet line (H$\beta -$) during the 2002 June and 2003 March runs at
McDonald Observatory and the apparent absence of jet lines during the
2003 October run at KPNO.  Upon closer inspection, we do find a
weak jet line blended with the H$\gamma$ line on night one of the KPNO
run.  We isolated the jet line for measurement by scaling
the night two spectrum
by the peak H$\gamma$ intensities and subtracting it from the
night one spectrum.  It is interesting
that the remaining nights do not show evidence of optical jet
lines.  Instances in which the jet lines disappear for days at a
time have been noted in the past and while not common, are not
exceedingly unusual \citep{mar84,gie02}.

\begin{figure}[p]
\begin{center}
\plotone{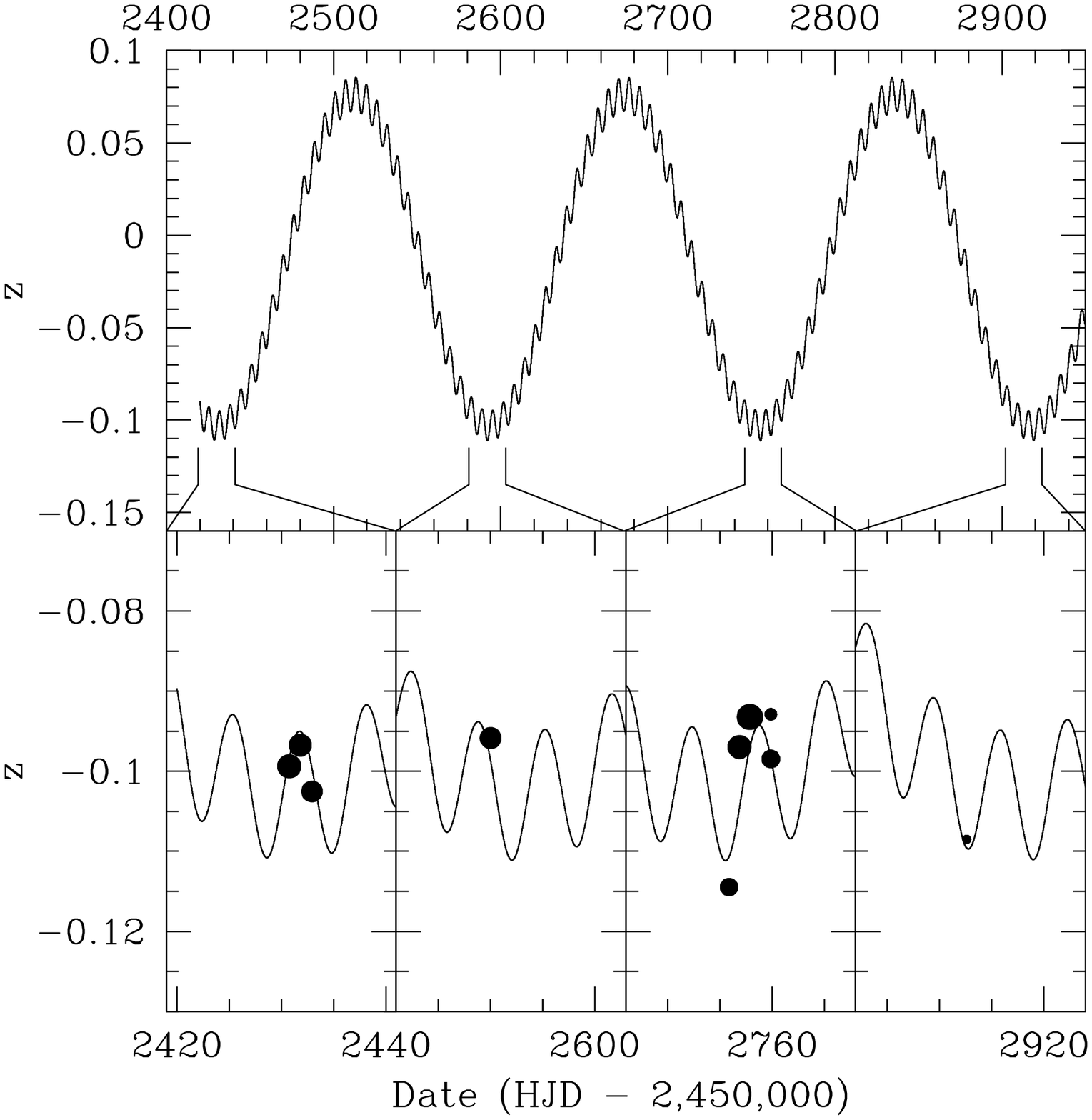}
\end{center}
\caption[Jet line observations and model] {The model of expected
approaching jet line velocity from \citet{gie02} along with our measured
jet features.  Point area is plotted proportional to equivalent width.
\label{jetfit}}
\end{figure}

Velocity changes in the jet lines from one night to the next
are due primarily to the ``nodding'' motion of the disk
caused by the orbiting mass donor \citep{kat82,gie02}.
\citet{gie02} produced a model fit of the disk nodding for
related jet motions of the H$\alpha +$ and H$\alpha -$ jet lines.
They treat the lines as individual ``bullets'' best measured
via Gaussian fits of distinct peaks.  We use this method
(see Table 2) and their model fit to illustrate the
Doppler shifts in Figure \ref{jetfit}.  The solid line in the figure
is the radial velocity curve from \citet{gie02} for the
H$\alpha -$ jet feature.  Each dot represents an observed jet
feature and the area of the dot is proportional to the measured equivalent
width of the feature.  In the last three nights of the 
2003 April spectra, the jet lines are blended with the diffuse
interstellar band at 4430 \AA .  To remove its effect,
the normalized spectrum from night one was set to an average value
of zero (by subtracting 1.0 from each point) and then subtracted
from the remaining three nights.  The equivalent widths were 
then measured for these nights from the subtracted spectra.

Two independent ``bullets'' were measured
in the last spectrum from 2003 April, giving two plotted points
in Figure \ref{jetfit}.  One of these is at the same velocity as
the feature measured in the previous spectrum and is likely
the remnant of that feature.
The tentative jet feature that we identified in the first spectrum
from the KPNO run is included in Figure \ref{jetfit} and
the good match between its velocity and the model
velocity curve suggests that this is indeed a jet emission line.
It is clear from Figure \ref{jetfit} that
the parameters found by \citet{gie02} are still appropriate when
applied to data for the H$\beta -$ jet obtained up to five years later.
Figure \ref{jetfit} also establishes that our spectra were indeed
taken around precessional phase $\Psi =0.0$
(minimum $z$ for the H$\beta -$ jet).

\begin{center}
\begin{deluxetable}{lrrrrrr}
\tablewidth{0pc}
\tablecolumns{7}
\tablecaption{Emission line equivalent widths and
jet-line radial velocities\label{jettab}}
\tablehead{
\colhead {Date}           & \colhead {$z$} &
\multicolumn{5}{c}{$W_\lambda$ (\AA)} \\ \cline{3-7}
\colhead {(HJD-2,450,000)}& \colhead {(Jet)}    & \colhead {Jet} &
\colhead {H$\gamma$} & \colhead {\ion{He}{1} $\lambda 4471$} &
\colhead {\ion{N}{3}/\ion{C}{3}} & \colhead {\ion{He}{2} $\lambda 4686$} }
\startdata \hline
2430.762\dotfill& $-0.099$ & $-18.1$  & $-16.8$  & $-3.3 $ & $-10.2$  & $-22.4$ \\
2431.823\dotfill& $-0.097$ & $-16.5$  & $-13.1$  & $-1.0 $ &$\phn-7.6$& $-17.1$ \\
2432.938\dotfill& $-0.103$ & $-14.4$  &$\phn-6.5$& $-0.7 $ &$\phn-7.7$& $-11.4$ \\
2590.037\dotfill& $-0.096$ & $-14.7$  & $-14.7$  & $-2.8 $ &$\phn-9.0$& $-16.2$ \\
2755.873\dotfill& $-0.115$ & $-10.5$  &$\phn-4.3$& $-0.4 $ &$\phn-5.5$& $\phn-6.7 $ \\
2756.873\dotfill& $-0.097$ & $-17.8$  &$\phn-7.4$& $-1.0 $ &$\phn-8.1$& $-12.0$ \\
2757.870\dotfill& $-0.093$ & $-21.5$  & $-12.2$  & $-2.3 $ & $-11.6$  & $-16.4$ \\
2759.878\dotfill& $-0.093$ &$\phn-5.0$&$\phn-4.5$& $-1.1 $ &$\phn-5.6$& $\phn-8.6 $ \\
2759.878\dotfill& $-0.098$ & $-10.8$  & \nodata  &\nodata  & \nodata  & \nodata\\
2912.695\dotfill& $-0.109$ &$\phn-2.6$&$\phn-6.3$& $-0.7 $ &$\phn-8.6$& $-10.4$ \\
2913.692\dotfill& \nodata  & \nodata  &$\phn-8.4$& $-0.9 $ &$\phn-8.4$& $-12.5$ \\
2914.697\dotfill& \nodata  & \nodata  & $-13.6$  & $-2.5 $ & $-11.7$  & $-20.2$ \\
2915.689\dotfill& \nodata  & \nodata  & $-12.0$  & $-1.8 $ & $-10.7$  & $-15.8$ \\
\enddata
\end{deluxetable}
\end{center}

\subsection{Stationary Emission Lines}

Because our spectra are not flux calibrated we need a different method
to confirm that we were indeed observing the system during
primary eclipse.  \citet{gie02} and GHM03 point out that many
of the emission lines are expected to form
in a volume larger than that of the inner disk.  Lines of this
type would show a small decrease in absolute flux during eclipse.
On the other hand, emission lines formed in the inner regions of the
disk should also be eclipsed by the donor star and should
show a corresponding drop in absolute flux.

With our rectified continuum, the {\it apparent} behavior of these lines
would be different.  As the continuum light decreases, the ``wind''
lines would appear to increase in strength (relative to the continuum)
while the inner disk lines should remain roughly constant in
strength (again, relative to the continuum).  The H$\gamma$,
\ion{He}{1} $\lambda 4471$, and possibly \ion{He}{2}
$\lambda 4686$ emission lines
are expected to be ``wind'' lines.  By measuring the strength of
these lines relative to the continuum, we can test that our
observations did indeed take place during eclipse.
The origin of the \ion{N}{3}/\ion{C}{3} complex at
$\lambda 4650$ is less certain, though for completeness we measure it
as well.

\begin{figure}[p]
\begin{center}
\plotone{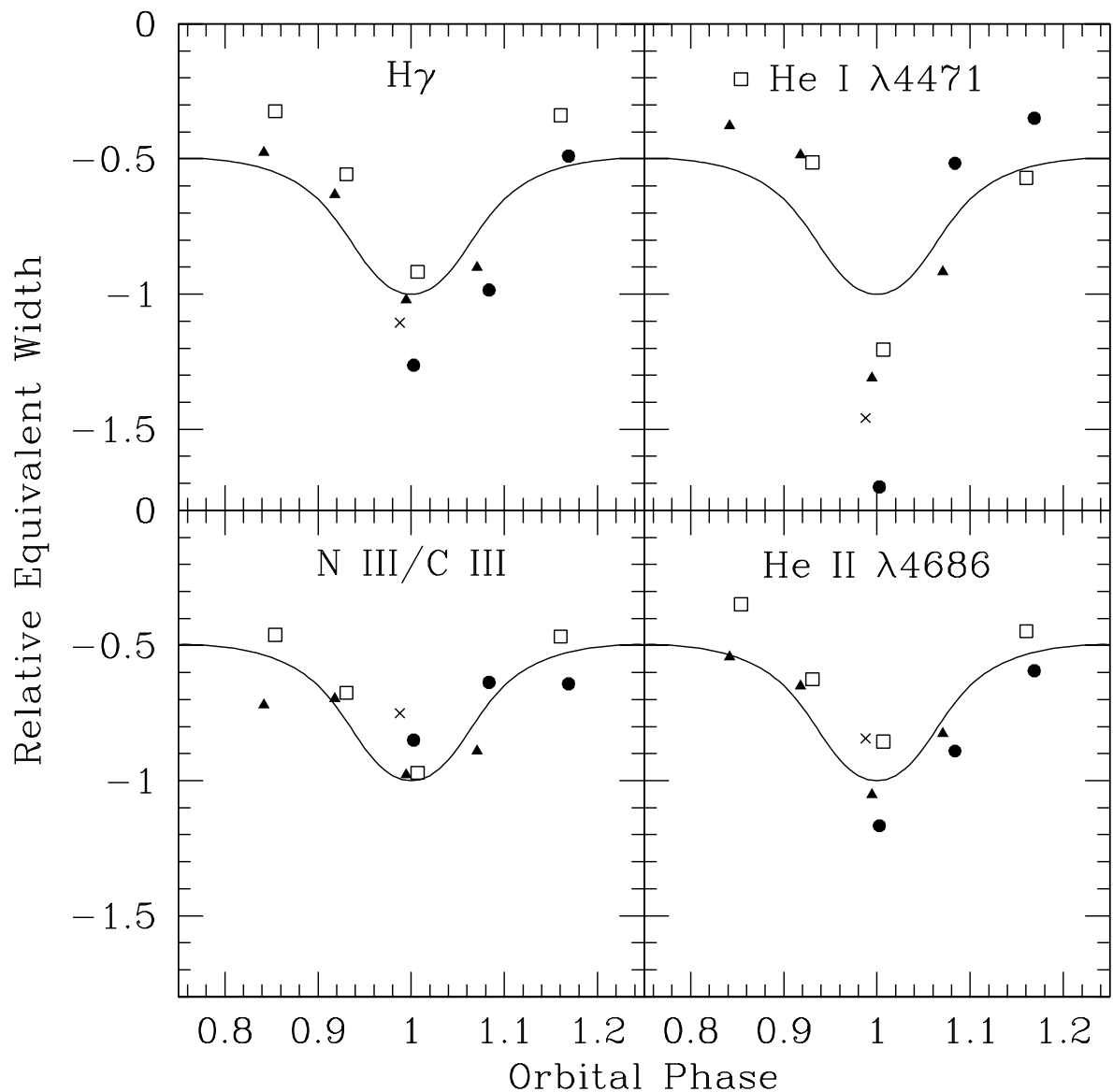}
\end{center}
\caption[Equivalent Width vs. Orbital Phase] {Equivalent width
measurements for the four major emission features seen in spectra
from all four observing runs plotted vs. orbital phase.  Filled
circles represent the 2002 June McDonald run, x's the 2002
November McDonald observation, open squares the 2003 April
McDonald run, and filled triangles the 2003 October KPNO run.
The solid line is a schematic representation of the expected
variations based upon the
$B$-band eclipse light curve, derived from \citet{gor97,gor98}
\label{em}}
\end{figure}

Our measured values for the equivalent width of the major emission
lines are given in Table \ref{jettab}.  The average errors for
measurements in the table are $\sigma _z \approx 0.001$ and
$\sigma _{W_\lambda}\approx 0.1$ \AA .
Figure \ref{em} shows the relative strengths of each of the four
emission features for each run.  We also show a
schematic representation of the
expected variations through eclipse of a constant flux source in
our continuum-rectified versions of the spectra.  The $B$-band flux
variation was estimated by a spline fit through the average magnitude
in each 0.05 phase bin of the $V$-band light curve for $\Psi = 0.0$
from \citet[see their Fig.~7(c)]{gor98} adjusted for the $B-V$
color variation reported by \citet[see their Fig.~3]{gor97}.
The strengths of the measured emission features are normalized
in each case to best match the schematic eclipse curve.
The H$\gamma$, \ion{He}{1} $\lambda 4471$, and \ion{He}{2}
$\lambda 4686$ lines all show a clear eclipse effect centered
on the night in each run with calculated orbital phase closest
to $\phi=0.0$ and they all appear to match the general behavior of
the eclipse light curve.  This confirms that our ephemeris
was correct and that our spectra were taken during primary
eclipse.

We also see from Figure \ref{em} that the \ion{N}{3}/\ion{C}{3}
complex does not
show such clear variations.  The 2003 April McDonald data ({\it open
squares}) show behavior very much like that seen in the other wind lines.
During the remaining runs, however, this feature shows only a small
modulation near $\phi=0.0$.  It is likely that the \ion{N}{3}/\ion{C}{3}
lines are confined to more interior regions of the disk and that during
the 2003 April run the \ion{N}{3}/\ion{C}{3} emitting region had expanded.

\section{The Absorption Spectrum of SS~433}

In order to best identify the absorption spectrum of the mass donor
star, we identified a portion of the continuum that was free of large
emission lines.  The best available region in our spectra 
extends from 4510--4625\AA~(Fig. \ref{mespec}).
This range avoids the interstellar absorption at 4502\AA , emission from
\ion{He}{1} $\lambda 4471$, and the \ion{N}{3}/\ion{C}{3}
emission complex which begins near 4630\AA .  Since the 2003
October KPNO spectra have the highest S/N ratio and resolution,
we compared the mid-eclipse spectrum by eye with spectra
of the comparison stars that we observed.  Because GHM03 had
not found a good match with the spectra of an O or B-type star,
but had identified possible
matches with an A-type supergiant, we obtained spectra during the KPNO
run of supergiants of types A4~Iab, F5~I, and G8~Iab.  During the previous
McDonald runs we obtained spectra of stars of type B1~Iab, B5~Ib,
A0~Ib, A7~Ib, and F0~Iab.

\begin{figure*}[p]
\begin{center}
\epsscale{0.9}
\plotone{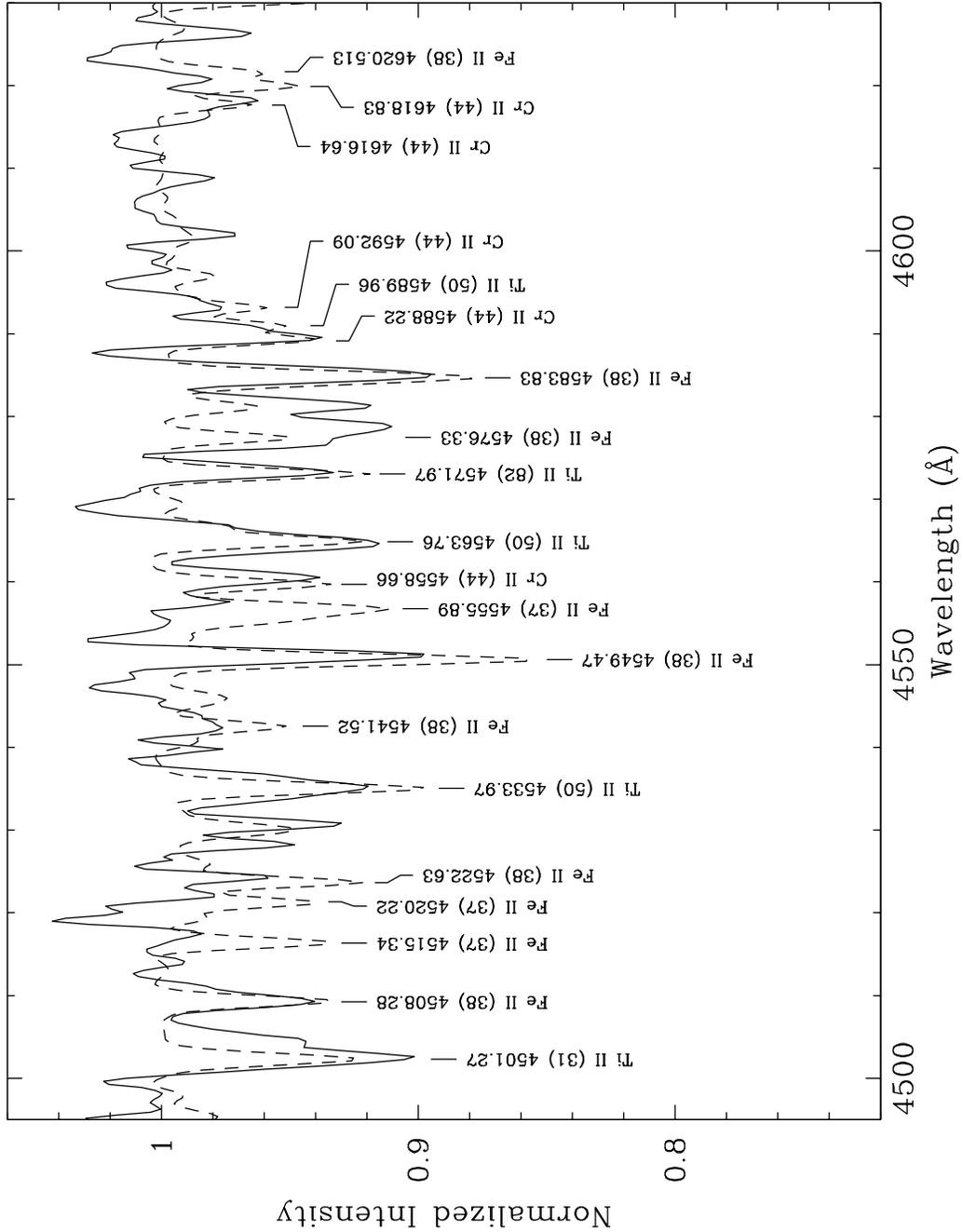}
\end{center}
\caption[SS~433 continuum and A4~Iab overplot] {An overplot of the
continuum region of SS~433 from night three (near mid-eclipse) of the
2003 October KPNO run ({\it solid line}) and the A4~Iab star HD~9233
({\it dashed line}).  The spectrum of HD~9233 has been Doppler-shifted
and intensity scaled by 36\% to match the spectrum of SS~433 (see text).
\label{overplot}}
\end{figure*}

The A4~Iab spectrum (of HD~9233) appeared to have a significant number
of features in common with the absorption line patterns in
the chosen continuum region of SS~433.
The spectra were boxcar smoothed by 3 pixels to increase
the S/N ratio and
again compared by eye.  The apparent correlation between the two
spectra became more obvious.  The smoothed spectra of SS~433
({\it solid line}) and HD~9233, the A4~Iab star ({\it dashed line}),
are shown overplotted in
Figure \ref{overplot}.  The A4~Iab spectrum has been Doppler-shifted
according to the relative velocity shift found via cross-correlation
functions (CCFs), as described below.  We have also applied an intensity
scaling factor to the A star spectrum to match the equivalent
widths of the absorption features in SS~433.  Because of the 
continuum light contribution of the disk/wind, the spectrum of
the mass donor will be veiled.  The depth of the absorption
features will be smaller relative to the continuum than in the
case of HD~9233 where the A star is the only continuum source.
The applied correction
factor is $0.36\pm 0.07$, so that the lines in the
A star spectrum are plotted with $36\%$ of
their observed strength.  If the absorption spectrum belongs to the
donor star then this comparison tells us that the mass donor in
SS~433 contributes $\approx 36\%$ of the total light at mid-eclipse.

The spectrum of SS~433 in Figure \ref{overplot} has been highly
rectified to remove any low-order variations in the spectra
(compare Fig. \ref{overplot} with Fig. \ref{mespec}) since
we are attempting to match only high-order variations---the narrow
absorption lines.  We have labeled a number of the strongest
absorption lines with their element, species, wavelength,
and multiplet number.  It is interesting to note that the lines
which seem to match least well correspond to the Fe II (37)
multiplet, yet the Fe II (38) multiplet lines are all well matched.
In addition to that of the A4~Iab star, the spectrum of the 
A7~Ib supergiant HD~148743 also
appeared to fit the spectrum of SS~433 quite well.  The lower
resolution of the McDonald spectrum of HD~148743, however, did not
provide as useful a comparison to the higher S/N ratio spectra
from the 2003 October KPNO run.
None of the remaining spectral standard stars that we observed provided
as good a match with the SS~433 spectrum.  

In order to quantify the correlation between the SS~433 spectrum and
the A4~Iab spectrum, we cross-correlated all of the SS~433 spectra
with the intensity scaled spectrum of HD~9233.  The A4~Iab spectrum
was smoothed to the resolution of the spectra for each of the
McDonald runs using a Gaussian smoothing algorithm.

All four of the
2003 October KPNO spectra produced CCFs which showed a strong
correlation peak.  The single McDonald spectrum from the
2002 November run was simply too
noisy for a good cross-correlation.  The spectra from the two
remaining McDonald
runs of 2002 June and 2003 April did not provide conclusive
correlations due to low S/N ratios.  Peaks were present in the
CCFs for these runs near the expected velocity shifts, but they
were not clearly significant when compared to the ``noise'' in the CCF.
By overplotting the A4 spectrum, we found that
a number of major absorption lines did appear in common.
While this alignment of features was
significant enough in the 2002 June spectra for GHM03 to produce
a tentative classification, as mentioned previously, the S/N ratio
was simply not high enough to produce a clear correlation.

A comparison of the strengths of the Ti II, Cr II, and Fe II absorption
lines in both spectra shows that the relative depths of absorption
features in SS~433 are very similar to that in HD~9233.  The
S/N ratio severely limits this analysis, but we can say that
the overall pattern of lines visible in SS~433 appears much
more like the A4 spectrum than the F0 or A0 McDonald spectra.
Therefore, considering the cross-correlation results and visual
inspection, we tentatively classify the absorption spectrum
as A3-7~I, with a corresponding effective temperature
of $T_{\rm eff}=8500 \pm 1000$~K \citep{ven95}.

\begin{figure}[p]
\begin{center}
\plotone{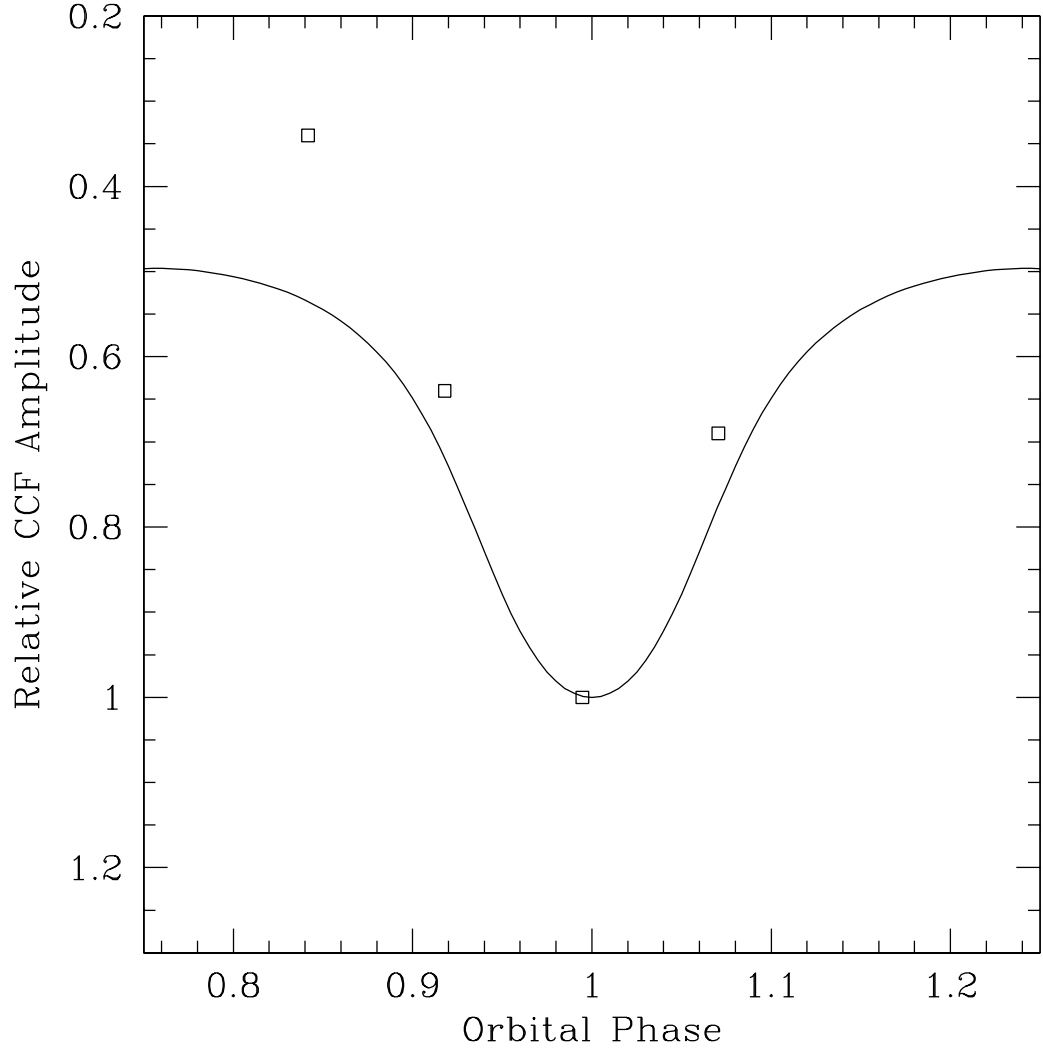}
\end{center}
\caption[CCF Amplitude vs. Orbital Phase] {Cross-correlation
amplitudes for the absorption features of the four spectra
from the 2003 October KPNO run ({\it squares}).
Also shown is the predicted variation for a constant flux source
({\it solid line}) based upon the schematic $B$-band light curve
(see text).
\label{absew}}
\end{figure}

We expect that absorption lines from
the mass donor should
increase in strength relative to the continuum as SS~433 goes
into eclipse and a larger fraction of the observed light is coming
from the donor star.  This should be followed by a decrease in
relative strength during egress of the eclipse.  This
trend should appear in the CCFs as an increase, then decrease in
the amplitude of the peak.  The relative CCF peak amplitudes for the
KPNO spectra are measured to be 0.34, 0.64, 1.00, and 0.69 from
first to fourth night, respectively.  The approximate error on the
amplitudes is $\pm0.08$.  We would expect that the
strength of absorption
features from the mass donor and emission lines originating in the
extended disk wind should behave similarly during eclipse.
Figure \ref{absew} shows the absorption line cross-correlation
amplitude variation plotted similarly to the emission line equivalent
widths in Figure \ref{em}.  Also shown is the prediction of the
variation for a constant flux source based upon the estimated
$B$-band light curve ({\it solid line}).
The differences in the observed and predicted variation are
qualitatively what would be
expected if the donor star is being obscured as it descends into
an extended disk.

The CCFs also provide us with measurements of the radial velocities
of the absorption features relative to those of HD~9233.  To convert
these to absolute radial velocities, we used Gaussian fits to measure
velocities for 13 lines in the spectrum of HD~9233.  The average of
these lines produced a radial velocity measure for HD~9233 of
$-34\pm2$ km s$^{-1}$, which we added to the relative
velocities of the CCFs.
Our values for $V_r$ are given in Table \ref{vrad} for the four
2003 October KPNO spectra.
\begin{center}
\begin{deluxetable}{lcc}
\tablewidth{0pc}
\tablecolumns{3}
\tablecaption{SS~433 absorption line radial velocities\label{vrad}}
\tablehead{
\colhead {Date}           & \colhead {$V_r$}     & \colhead {error} \\
\colhead {(HJD-2,450,000)}& \colhead {(km s$^{-1}$)}&\colhead {(km s$^{-1}$)} }
\startdata \hline
2912.695\dotfill& 25 & $ 6$ \\
2913.692\dotfill& 40 & $ 5$ \\
2914.697\dotfill& 68 & $ 4$ \\
2915.689\dotfill& 76 & $ 7$ \\
\enddata
\end{deluxetable}
\end{center}

\begin{figure}[p]
\begin{center}
\plotone{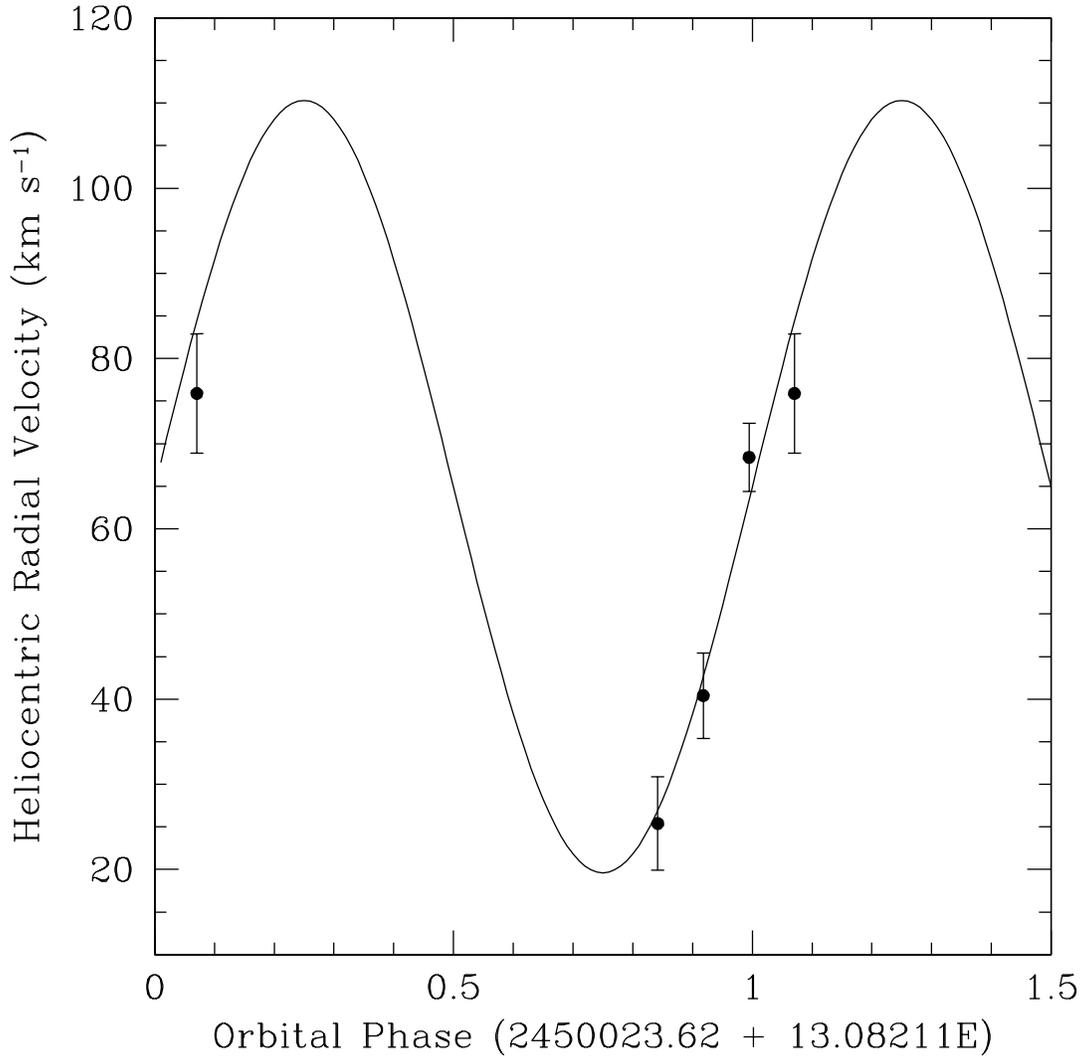}
\end{center}
\caption[Sine curve fit to absorption line radial velocities]
{The best-fit sine curve to the four radial velocities obtained
via cross-correlation fitting of the 2003 October KPNO spectra.
All absorption radial velocity values are in a heliocentric
rest frame.\label{abs_fit}}
\end{figure}

Although constructing a radial velocity curve from only four
points is difficult, most of the
orbital parameters for SS~433 are already known.  With the
light curve ephemeris of
\citet{gor98}, as used above, the only remaining parameters for the
radial velocity curve of the mass donor are systemic velocity, $\gamma$,
and semi-amplitude, $K_O$.  We use subscripts $O$ and $X$ to
distinguish the motion of the optical star and the compact X-ray source.
Using a non-linear least squares routine,
we fit our four radial velocities with a sine curve, assuming zero
orbital eccentricity \citep{fab90}.
The resulting fit gives $\gamma=65\pm3$ km s$^{-1}$ and
$K_O=45\pm6$ km s$^{-1}$.  Figure \ref{abs_fit} shows our four
radial velocity points and the best-fit sine curve.

\citet{fab90} give a semi-amplitude for the compact object and disk
of $K_X=175\pm20$ km s$^{-1}$ and \citet{gie02} give 
$K_X=162\pm29$ km s$^{-1}$.
Taking the average of these two, $K_X=168\pm18$ km s$^{-1}$, we arrive at
a mass ratio, $q=M_X/M_O=K_O/K_X=0.27\pm0.05$.
This value of $q$ is interestingly near the convergence of the
two mass ratio predictions of \citet{ant87} and \citet{ant92}.
From optical light curve modeling, \citet{ant87} arrive at
a mass ratio lower limit, $q\geq 0.25$.  However, from X-ray
eclipse models, \citet{ant92} arrive at a range of possible
mass ratios, $q=$~0.15--0.25.  Our value of $q=0.27$ is consistent
with both suggestions, falling
at the intersection of these two models.

In addition to velocities, we can also model the projected rotational
velocity $v_{rot}\sin i$ of the star.
The modeling was done by comparing theoretical line
broadening functions to our observed spectra.
The instrumental broadening, found from our comparison spectra,
is 22.5 km s$^{-1}$.  A Gaussian with this width ($\sigma$),
was broadened assuming a spherical star with a linear limb-darkening
coefficient of 0.57 \citep{wad85}.  A grid of theoretical profiles was
computed for a range of $v_{rot}\sin i$ values and a $\chi^2$
minimization was used to determine the best fit to the data.
For the mid-eclipse spectrum from KPNO, which has the deepest
lines, the average $v_{rot}\sin i$ of the eight strongest
lines is $80\pm 20$ km s$^{-1}$.  The same lines for the
A4~Iab star HD~9233 give a $v_{rot}\sin i$ of $39\pm 12$ km s$^{-1}$
which is at the upper end of the range found for A supergiants 
by \citet{ven95}.
Using a system inclination for SS~433 of $78\fdg8$ \citep{mar89}
and assuming $i_{spin}=i_{orbit}$
gives a rotational velocity, $v_{rot}\approx82$ km s$^{-1}$.
The cause of the disk precession is unknown, but if precession
of the rotating donor star is the cause, then the spin and
orbital axes may be misaligned.

\section{Discussion}

From the previous section, we see that the radial velocity curve
of the absorption lines is in anti-phase to that of the disk, the
resulting mass ratio is within the bounds of existing models of the
system, and the magnitude of the CCF amplitude variations with
orbital phase is what we might expect if the features originate
from the mass donor star.  Each of these results points to an
origin of the A-type absorption spectrum in the mass donor
star.  If we assume that these lines do originate in the mass
donor, then we can make several further calculations.

We may determine the component masses using the equation
$$M_{O,X}=(1.0361\times10^{-7})(\sin i)^{-3}(K_O+K_X)^2K_{X,O}
P~{M}_\odot,$$
where the semi-amplitudes are in units of km~s$^{-1}$
and the period is in days.
Our value of $K_O=45\pm6$ km~s$^{-1}$ from the above and the
average value of $K_X=168\pm18$ km~s$^{-1}$, along with $i=78\fdg8$
from kinematical models of the jets \citep{mar89},
result in a mass for the donor star of $M_O=10.9\pm3.1~M_\odot$
and a mass of the compact object/disk of $M_X=2.9\pm0.7~M_\odot$.
Depending on the mass in the disk, our calculated
mass may support either a black hole or neutron star as
the compact companion.  Predictions vary widely regarding the
mass of the disk.  \citet{col02} suggest an extremely massive
disk, though their system parameters are significantly different
from ours.  On the other hand, \citet*{kin00} predict a
very low mass disk on the basis of evolutionary models.
\citet{kin00} also suggest that for a mass donor with
$M_O\gtrsim 5 M_\odot$ the companion is likely to be a black
hole.  Observations indicate that all neutron stars have a
mass close to $1.35 M_\odot$ \citep{tho99}, with the possible
exception of Vela X-1 for which a mass near $1.9 M_\odot$ has
been reported \citep{bar01,qua03}.
In light of these results and our calculated mass,
and in the absence of a massive disk,
we suggest that the compact companion in SS~433 is
a black hole rather than a neutron star.

With these masses we can calculate the binary separation
and Roche lobe radius for the mass donor.  The resulting
binary separation from Kepler's third law is $a=0.26\pm 0.02$ AU
$=56\pm 4 R_\odot$.  The volume Roche lobe radius
for the mass donor star can be found by \citep{egg83}
$$R_L=\frac{0.49q^{-2/3}a}{0.6q^{-2/3}+\ln (1+q^{-1/3})}$$
which gives $R_L=28\pm 2 R_\odot$.
This value is approximately
the expected radius for an A type supergiant \citep{ven95}.
The resulting surface gravity is $\log g = 2.59\pm 0.14$.

We can use this value for the Roche lobe radius and the orbital
period to derive an expected $v_{rot}$ for the mass donor if
it is synchronously rotating.  Assuming synchronous
rotation may not be valid for this system, as described below,
but it gives us a quantitative comparison to our measured
$v_{rot}\sin i$.  The resulting value is $v_{rot}=108\pm 7$
km s$^{-1}$, which is somewhat larger than our estimated value
of $82\pm 20$ km s$^{-1}$.

A possible explanation for the difference in observed $v_{rot}$
and expected synchronous $v_{rot}$ comes from the SS~433
evolutionary scenario of \citet{kin00}.  They suggest that
SS~433 is undergoing a period of rapid evolution as the mass
donor crosses the Hertzsprung gap.  This leads to extremely
high mass transfer rates.  Under these conditions, synchronous
rotation is not required and is possibly even unlikely.
An alternative explanation is that the spin axis of the mass donor
is not aligned with the orbital axis of the system.  This type
of misalignment may be possible as the result of an asymmetry
in the supernova that created the compact companion \citep{bra95}.

The scaling of the HD~9233 spectrum to match the mid-eclipse
SS433 spectrum from the 2003 October KPNO run also provides
information regarding the origin of the absorption features.
The 0.36 scaling factor means that, if these
lines do originate in the mass donor, the accretion disk
still contributes well over half of the light of the system
at mid-eclipse.  This is consistent with
the findings of \citet{gor98}.  They find a 0.41 mag precessional
variation in the central eclipse light curve.  
We must point out though, that our scaling factor
was found assuming a mass donor of luminosity type I and our spectrum
does not provide sufficient luminosity diagnostics to confirm
a luminosity type unambiguously.  If the mass donor is
of luminosity type II or higher, then the $36\%$ scaling
factor would be too low because the metal lines are weaker
in higher gravity stars.
However, the large mass transfer rate in the system probably
indicates that the mass donor fills its Roche lobe, and the
mass donor Roche lobe radius found above is consistent with
a luminosity type I \citep{ven95}.

We can compare our estimated scaling factor with that
expected from observed light curves of SS~433.
If we assume that the donor star is totally eclipsed by the
disk at $\phi =0.5$, then we can use the light
curve of \citet{gor98} to estimate the flux ratio as
$$\frac{F_\star}{F(\phi =0.00)}=\frac{F(\phi =0.25)-F(\phi =0.50)}
{F(\phi =0.00)}=0.29\pm0.08.$$
This estimate comes from a comparison of the $V$-band light curves of
\citet{gor98} and \citet{kem86}.  The result matches within errors
with our measured scaling factor of $0.36\pm 0.07$.
Note that \citet{gor97} find that the system is slightly
redder during mid-eclipse, which implies that the mid-eclipse
flux is relatively lower in the $B$ band compared to the $V$ band.
Thus, the predicted flux contribution of the mass donor in the
$B$ band may be slightly larger than given above.
The donor star flux contribution may also be larger if the
donor star is not completely eclipsed at $\phi =0.50$.

Regardless of the source of the absorption lines, as long as
it is within the system, we have the systemic radial
velocity from fitting the absorption line radial velocity
curve.  Several distance determinations to SS~433 have arrived
at values from 3.0 to 4.85 kpc \citep[e.g.][]{ver93,dub98,sti02}.
Differential galactic rotation curves give the expected
system velocity for these distances as 32 and 59 km s$^{-1}$,
respectively, for an object sationary with respect to its
local standard of rest \citep{ber01}.
Our value, $\gamma = 65\pm3$ km s$^{-1}$
falls very near the upper end of this range.  One must
be careful in comparing these two values too closely however
since the past supernova in SS433 which resulted
in the compact companion may have given the system a
runaway velocity \citep{bra95}.
The final $\gamma$ value may therefore not be directly comparable
to the results from the differential galactic rotation curve.

The discovery of a mid-A type supergiant spectrum in SS~433 does not
necessarily require that the mass donor itself is a mid-A supergiant.
One alternative is that we are seeing a ``shell'' spectrum.  The
Be star Pleione, for example, shows a narrow line spectrum
on top of the broad B star stellar spectrum due to the
projection of circumstellar disk gas against the visible
hemisphere of the star\citep{bal80}.
This ``shell'' spectrum closely resembles that of an early A supergiant.
Is it possible then that what we are seeing is some form of shell
spectrum from SS~433?  We doubt this origin for several reasons.
First, since our CCFs show a variation in amplitude, with the lines
appearing stronger nearest mid-eclipse, the shell spectrum must
be physically associated with the mass donor.  In other words,
the absorption taking place is absorption of continuum light
from the mass donor.  Likely this would
mean that we only see the shell spectrum from a region that is
shadowed from the high flux of the inner disk by the mass donor.
The density in this region would have to be high to produce the
observed A type spectrum.  Moreover, the strength of the lines
indicates a stellar flux contribution that matches that
expected from the light curve, as shown above.  So not only is
the light being absorbed
continuum light from the star, but the absorption must be
occurring over most or all of the stellar disk.  This inferred
combination of high opacity and full disk coverage suggests that
the lines have a photospheric rather than shell origin.
Second, the radial velocity variations we observe mean that any
``shell'' source must also be orbiting within the system near
the donor.  This is somewhat consistent with the idea of
a shadowed region, since the shadow would move with the mass donor.
The problem here is that the gas in the shadowed region is unlikely
to be gravitationally bound to the donor if the donor fills its
Roche lobe.  Thus, any ``shell'' gas component in this location
is likely transient and may not produce the Keplerian
motion we observe.  Third, the $v_{rot}\sin i$ we
measure in the SS~433 spectrum is at the upper end of values
for A supergiants.  It is doubtful that a shell spectrum in this
region would
produce $v_{rot}\sin i$ values this high.  One certain constraint we can
place in the event of a shell spectrum is that the spectral
type of the mass donor must be earlier than A7.  Were the mass
donor of later spectral type the dense shell would be hotter than
the star, which is very unlikely.

It is interesting to note that \citet{cha04} have also obtained
blue spectroscopy of SS~433, though at precessional phases
when the disk was nearly edge-on.  They show features in the
same spectral range as our results which also seem to indicate
the presence of a late A-type spectrum.  However, the velocities
that they arrive at using these absorption features are some
220 km s$^{-1}$ more negative than ours and do not show Keplerian
motion.  \citet{cha04} do
mention that the interpretation of their results will
be complicated if the mass donor is embedded in a dense
outflowing disk wind (likely occuring in the orbital plane).
This reiterates our reasoning for observing at precessional
phase zero, when the mass donor would be above such an
outflow rather than embedded in it.

Additionally, the absorption features found in the
\citet{cha04} spectra
appear to be quite different than we present here.
In addition to the very different radial velocities,
the features in their spectra are {\it much}
stronger and broader than those we have observed.  The type
of absorption they observe was also discussed by \citet{mar84b}.
He explains this absorption as a shell spectrum and notes
that the strength of the lines depends strongly on the
precessional phase, with the strongest lines appearing when
the disk is edge-on.  This is precisely when \citet{cha04}
made their observations.  \citet{mar84b} attributes these
lines to the accretion disk.  The lines we have observed
have very different characteristics.  They
appear to be much weaker than the lines in \citet{cha04},
vary through eclipse,
have a well-defined Keplerian velocity behavior, and have
the relative strength expected for lines from the photosphere
of the mass donor star, as described above.

\section{Conclusion}

We have shown that the mass donor in SS~433 is likely to be an
A3-7~I star.  A similar kind of absorption spectrum may be
produced by circumstellar gas, but we argue that the strength of
the spectrum and its orbital variation in intensity and Doppler
shift all indicate a photospheric origin.
Further observations with higher S/N ratio spectra would allow
us to classify more accurately the spectrum and calculate 
temperature, abundances, and surface gravity for the mass donor.
With additional radial velocity measurements to help constrain the fit,
more accurate masses would also be obtained which would
lead to a more secure identification of the compact star
as either a black hole or neutron star.

Our estimates of the mass ratio and masses fall within the values
predicted in the literature and begin to give us a more
definite understanding of the physical parameters of this
peculiar system.  The evolutionary scenario which best fits
our determined physical parameters is that of \citet{kin00},
in which the mass donor star is in a phase of rapid
evolution as it crosses the Hertzsprung gap and the compact
companion is a low mass black hole.

We have also shown that the orbital ephemeris of \citet{gor98}
and disk precession/nodding model parameters of \cite{gie02}
are still valid.

\acknowledgements
We thank the staffs of KPNO and McDonald Observatory for their
assistance in making these observations possible.  MAS would
like to thank T. Bogdanovic for assisting with the
observing run.
We are grateful to Bruce Margon for comments on an early
draft of this work.
Financial support was provided by the National Science Foundation
through grant AST$-0205297$ (DRG).
Institutional support has been provided from the GSU College
of Arts and Sciences and from the Research Program Enhancement
fund of the Board of Regents of the University System of
Georgia, administered through the GSU Office of the Vice
President for Research.  Additional funding was provided
through a Zaccheus Daniel Foundation Fellowship (MAS).


\end{document}